\documentclass[aps,prl,twocolumn,superscriptaddress,showpacs,preprintnumbers,amsmath,amssymb]{revtex4}
\usepackage{graphicx,color} 
\usepackage{dcolumn}  
\usepackage{amsmath}  
\usepackage{refmerge}

\graphicspath{{ps/}}

\begin{document}


\def\LR{\mathcal{R}}
\def\dE{{\Delta E}}
\def\mb{{M_{\rm bc}}}
\def\Dt{\Delta t}
\def\Dz{\Delta z}
\def\egcms{{E_\gamma^{\rm c.m.s.}}}
\newcommand{\dmd}{\Delta m_d}
\def\fol{f_{\rm ol}}
\def\fsig{f_{\rm sig}}
\def\fbb{f_{B\bbar}}

\newcommand{\fCP}{f_{\rm rec}}
\newcommand{\zCP}{z_{\rm rec}}
\newcommand{\tCP}{t_{\rm rec}}
\newcommand{\ftag}{f_{\rm tag}}
\newcommand{\ttag}{t_{\rm tag}}
\newcommand{\ztag}{z_{\rm tag}}
\newcommand{\cala}{{\mathcal A}}
\newcommand{\cals}{{\mathcal S}}
\def\rsigbkg{{\mathcal R}_{\rm s/b}}
\def\rsigbkgBH{{\mathcal R}_{\rm s/b}^{\rm BH}}
\def\Rsig{R_{\rm sig}}
\def\Rbb{R_{B\bbar}}
\def\Rbkg{R_{\qq}}
\def\calf{{\mathcal F}}
\def\taubz{{\tau_\bz}}
\def\taubp{{\tau_\bp}}
\def\tauc{{\tau_{\ks\pip\gamma}}}
\def\taun{{\tau_{\ks\piz\gamma}}}
\newcommand*{\fq}{\ensuremath{q}}

\def\bz{{B^0}}
\def\bzb{{\overline{B}{}^0}}
\def\bbar{{\overline{B}}}
\def\bp{{B^+}}
\def\bm{{B^-}}
\def\kz{{K^0}}
\def\ks{{K_S^0}}
\def\kp{{K^+}}
\def\km{{K^-}}
\def\pip{{\pi^+}}
\def\pim{{\pi^-}}
\def\piz{{\pi^0}}
\def\kstar{{K^\ast}}
\def\kstarz{{K^{\ast 0}}}
\def\kstarp{{K^{\ast +}}}
\def\kstarm{{K^{\ast -}}}
\def\ktwostar{{K_2^\ast}}
\def\ktwostarz{{K_2^{\ast 0}}}
\def\kstarpm{{K^{\ast\pm}}}
\def\jpsi{{J/\psi}}
\def\qq{q\bar{q}}

\def\GeV{\,{\rm GeV}}
\def\GeVc{\,{\rm GeV}/c}
\def\GeVcc{\,{\rm GeV}/c^2}
\def\MeV{\,{\rm MeV}}
\def\MeVcc{\,{\rm MeV}/c^2}

\newcommand{\BaBar}{{\sc B\hspace*{-0.2ex}a\hspace*{-0.2ex}B\hspace*{-0.2ex}a\hspace*{-0.2ex}r}}

\def\NBBsvdI{152\times 10^6}
\def\NBBofour{275\times 10^6}
\def\NBBofive{386\times 10^6}
\def\NBBosix{535\times 10^6}
\def\NBBadd{257\times 10^6}
\def\NBBsvdII{383\times 10^6}

\def\NkppimgammaCandInFit{52823}
\def\NkppimgammaCandInBox{5764}
\def\Nsigpm{3196.1}
\def\Nsigpme{70.4}
\def\SNpm{1.25}

\def\NKspipgammaCandInFit{9932}
\def\NKspipgammaCandInBox{1069}
\def\Nsigc{557.9}        
\def\Nsigce{28.4}	 
\def\NsigcMRI{433.7}	 
\def\NsigcMRIe{21.3}	 
\def\NsigcMRII{46.4}	 
\def\NsigcMRIIe{9.6}	 
\def\NsigcMRIII{79.5}	 
\def\NsigcMRIIIe{13.5}	 
\def\SNc{1.22}		 
\def\SNcMRI{3.35}	 
\def\SNcMRII{0.55}	 
\def\SNcMRIII{0.33}   	 

\def\NkspizgammaCandInFit{4078}
\def\NkspizgammaCandInBox{406} 

\def\Nsign{175.8}        
\def\Nsigne{17.8}	 
\def\NsignMRI{112.5}	 
\def\NsignMRIe{12.0}	 
\def\NsignMRII{28.7}	 
\def\NsignMRIIe{7.1}	 
\def\NsignMRIII{35.2}	 
\def\NsignMRIIIe{10.0}	 

\def\SNn{0.88}		 
\def\SNnMRI{1.91}	 
\def\SNnMRII{0.81}	 
\def\SNnMRIII{0.35}	 

\def\Nrarebn{22.6}
\def\Nrarebne{9.3}
\def\NrarebnMRI{7.7}
\def\NrarebnMRIe{4.2}
\def\NrarebnMRII{4.1}
\def\NrarebnMRIIe{2.4}
\def\NrarebnMRIII{5.3}
\def\NrarebnMRIIIe{5.2}

\newcommand{\asyme}[3]{{#1^{+#2}_{-#3}}}
\newcommand{\syme}[2]{{{#1}\pm {#2}}}
\newcommand{\asymasyme}[5]{{#1^{+#2}_{-#3}{}^{+#4}_{-#5}}}
\newcommand{\symasyme}[4]{{{#1}\pm {#2}{}^{+#3}_{-#4}}}
\newcommand{\asymsyme}[4]{{{#1}^{+#2}_{-#3}\pm {#4}}}
\newcommand{\symsyme}[3]{{{#1}\pm {#2}\pm {#3}}}
\newcommand{\asymasymeSS}[5]{{#1^{+#2}_{-#3}\mbox{(stat)}{}^{+#4}_{-#5}\mbox{(syst)}}}
\newcommand{\symasymeSS}[4]{{{#1}\pm {#2}\mbox{(stat)}{}^{+#3}_{-#4}\mbox{(syst)}}}
\newcommand{\asymsymeSS}[4]{{{#1}^{+#2}_{-#3}\mbox{(stat)}\pm {#4}\mbox{(syst)}}}
\newcommand{\symsymeSS}[3]{{{#1}\pm {#2}\mbox{(stat)}\pm {#3}\mbox{(syst)}}}

\def\rarebzlife{\syme{1.356}{0.045}}
\def\rarebclife{\syme{1.505}{0.029}}


\def\SkstarzgmVallast{-0.79}
\def\SkstarzgmStatplast{0.63}
\def\SkstarzgmStatnlast{0.50}
\def\SkstarzgmSystlast{0.10}
\def\SkstarzgmResultlast{\asymsyme{\SkstarzgmVallast}{\SkstarzgmStatplast}{\SkstarzgmStatnlast}{\SkstarzgmSystlast}}

\def\SkspizgmVallast{-0.58}
\def\SkspizgmStatplast{0.46}
\def\SkspizgmStatnlast{0.38}
\def\SkspizgmSystlast{0.11}
\def\SkspizgmResultlast{\asymsyme{\SkspizgmVallast}{\SkspizgmStatplast}{\SkspizgmStatnlast}{\SkspizgmSystlast}}
\def\AkspizgmVallast{+0.03}
\def\AkspizgmStatlast{0.34}
\def\AkspizgmSystlast{0.11}
\def\AkspizgmResultlast{\symsyme{\AkspizgmVallast}{\AkspizgmStatlast}{\AkspizgmSystlast}}

%
\def\controllife{\syme{1.49}{0.10}}

\def\signallife{\asyme{1.53}{0.19}{0.17}}


\def\controlSin{\syme{0.20}{0.18}}
\def\controlCos{\syme{0.05}{0.11}}


\def\SkspizgmVal{-0.10}
\def\SkspizgmStat{0.31}
\def\SkspizgmSyst{0.07}
\def\AkspizgmVal{-0.20}
\def\AkspizgmStat{0.20}
\def\AkspizgmSyst{0.06}
\def\SkspizgmResult{\symsyme{\SkspizgmVal}{\SkspizgmStat}{\SkspizgmSyst}}
\def\SkspizgmResultSS{\symsymeSS{\SkspizgmVal}{\SkspizgmStat}{\SkspizgmSyst}}
\def\AkspizgmResult{\symsyme{\AkspizgmVal}{\AkspizgmStat}{\AkspizgmSyst}}
\def\AkspizgmResultSS{\symsymeSS{\AkspizgmVal}{\AkspizgmStat}{\AkspizgmSyst}}

\def\SI{-0.32}
\def\SStatIp{0.36}
\def\SStatIm{0.33}
\def\SSystI{0.05}
\def\SIResult{\asymsyme{\SI}{\SStatIp}{\SStatIm}{\SSystI}}

\def\AI{-0.20}
\def\AStatI{0.24}
\def\ASystI{0.05}
\def\AIResult{\symsyme{\AI}{\AStatI}{\ASystI}}

\def\SMRIResult{\asymsyme{\SI}{\SStatIp}{\SStatIm}{\SSystI}}
\def\SMRIResultSS{\asymsymeSS{\SI}{\SStatIp}{\SStatIm}{\SSystI}}
\def\AMRIResult{\symsyme{\AI}{\AStatI}{\ASystI}}
\def\AMRIResultSS{\symsymeSS{\AI}{\AStatI}{\ASystI}}

\def\SII{2.21}
\def\SStatIIp{+0.76}
\def\SStatIIm{-0.89}
\def\SIIResult{\asyme{\SII}{\SStatIIp}{\SStatIIm}}
\def\AII{-0.67}
\def\AStatII{0.54}
\def\AIIResult{\asyme{\AII}{\AStatII}}

\def\SIII{-0.75}
\def\SStatIIIp{+0.76}
\def\SStatIIIm{-0.54}
\def\SIIIResult{\asyme{\SIII}{\SStatIIIp}{\SStatIIIm}}
\def\AIII{+0.11}
\def\AStatIIIp{+0.50}
\def\AStatIIIm{-0.54}
\def\AIIIResult{\asyme{\AIII}{\AStatIIIp}{\AStatIIIm}}


\def\Snon{+0.50}
\def\SStatnon{0.61}
\def\SSystnon{0.29}
\def\SnonResult{\symsyme{\Snon}{\SStatnon}{\SSystnon}}
\def\SnonResultSS{\symsymeSS{\Snon}{\SStatnon}{\SSystnon}}
\def\Anon{-0.20}
\def\AStatnon{0.37}
\def\ASystnon{0.13}
\def\AnonResult{\symsyme{\Anon}{\AStatnon}{\ASystnon}}
\def\AnonResultSS{\symsymeSS{\Anon}{\AStatnon}{\ASystnon}}

\preprint{\vbox{ \hbox{   }
    \hbox{BELLE-CONF-0675}
    \hbox{hep-ex/0608017}
  }
} 

\title{ \quad\\[0.5cm]  Time-Dependent {\boldmath $CP$}
Asymmetries in $\bz\to\ks\piz\gamma$ Transitions\\
}
\date{\today}

\affiliation{Budker Institute of Nuclear Physics, Novosibirsk}
\affiliation{Chiba University, Chiba}
\affiliation{Chonnam National University, Kwangju}
\affiliation{University of Cincinnati, Cincinnati, Ohio 45221}
\affiliation{The Graduate University for Advanced Studies, Hayama, Japan} 
\affiliation{University of Hawaii, Honolulu, Hawaii 96822}
\affiliation{High Energy Accelerator Research Organization (KEK), Tsukuba}
\affiliation{University of Illinois at Urbana-Champaign, Urbana, Illinois 61801}
\affiliation{Institute of High Energy Physics, Chinese Academy of Sciences, Beijing}
\affiliation{Institute of High Energy Physics, Vienna}
\affiliation{Institute of High Energy Physics, Protvino}
\affiliation{Institute for Theoretical and Experimental Physics, Moscow}
\affiliation{J. Stefan Institute, Ljubljana}
\affiliation{Kanagawa University, Yokohama}
\affiliation{Korea University, Seoul}
\affiliation{Kyungpook National University, Taegu}
\affiliation{Swiss Federal Institute of Technology of Lausanne, EPFL, Lausanne}
\affiliation{University of Maribor, Maribor}
\affiliation{University of Melbourne, Victoria}
\affiliation{Nagoya University, Nagoya}
\affiliation{Nara Women's University, Nara}
\affiliation{National Central University, Chung-li}
\affiliation{National United University, Miao Li}
\affiliation{Department of Physics, National Taiwan University, Taipei}
\affiliation{H. Niewodniczanski Institute of Nuclear Physics, Krakow}
\affiliation{Nippon Dental University, Niigata}
\affiliation{Niigata University, Niigata}
\affiliation{Osaka City University, Osaka}
\affiliation{Osaka University, Osaka}
\affiliation{Panjab University, Chandigarh}
\affiliation{Peking University, Beijing}
\affiliation{Princeton University, Princeton, New Jersey 08544}
\affiliation{RIKEN BNL Research Center, Upton, New York 11973}
\affiliation{Saga University, Saga}
\affiliation{University of Science and Technology of China, Hefei}
\affiliation{Seoul National University, Seoul}
\affiliation{Shinshu University, Nagano}
\affiliation{Sungkyunkwan University, Suwon}
\affiliation{University of Sydney, Sydney NSW}
\affiliation{Toho University, Funabashi}
\affiliation{Tohoku Gakuin University, Tagajo}
\affiliation{Tohoku University, Sendai}
\affiliation{Department of Physics, University of Tokyo, Tokyo}
\affiliation{Tokyo Institute of Technology, Tokyo}
\affiliation{Tokyo Metropolitan University, Tokyo}
\affiliation{Tokyo University of Agriculture and Technology, Tokyo}
\affiliation{Virginia Polytechnic Institute and State University, Blacksburg, Virginia 24061}
\affiliation{Yonsei University, Seoul}
  \author{Y.~Ushiroda}\affiliation{High Energy Accelerator Research Organization (KEK), Tsukuba} 
  \author{K.~Sumisawa}\affiliation{High Energy Accelerator Research Organization (KEK), Tsukuba} 
  \author{K.~Abe}\affiliation{High Energy Accelerator Research Organization (KEK), Tsukuba} 
  \author{K.~Abe}\affiliation{Tohoku Gakuin University, Tagajo} 
  \author{I.~Adachi}\affiliation{High Energy Accelerator Research Organization (KEK), Tsukuba} 
  \author{H.~Aihara}\affiliation{Department of Physics, University of Tokyo, Tokyo} 
  \author{D.~Anipko}\affiliation{Budker Institute of Nuclear Physics, Novosibirsk} 
  \author{K.~Arinstein}\affiliation{Budker Institute of Nuclear Physics, Novosibirsk} 
  \author{A.~M.~Bakich}\affiliation{University of Sydney, Sydney NSW} 
  \author{E.~Barberio}\affiliation{University of Melbourne, Victoria} 
% ...
\collaboration{The Belle Collaboration}


\begin{abstract}
 We report measurements of $CP$ violation parameters in
 $\bz\to\ks\piz\gamma$ transitions based on a data sample
of $\NBBosix B\bbar$ pairs
collected
 with the Belle detector at the KEKB asymmetric-energy $e^+ e^-$
 collider. One neutral $B$ meson is fully reconstructed in the
 $\bz\to\ks\piz\gamma$ mode.
 The flavor of the accompanying $B$ meson is identified from its
 decay products.
 We obtain time-dependent and direct $CP$ violation parameters
 $\cals$ and $\cala$ for a $\ks\piz$ invariant mass up to $1.8\GeVcc$
 as $\cals_{\ks\piz\gamma}=\SkspizgmResult$ and
 $\cala_{\ks\piz\gamma}=\AkspizgmResult$.
 For a $\ks\piz$ invariant mass near the $\kstarz(892)$ resonance, we obtain
 $\cals_{\kstarz\gamma}=-0.32^{+0.36}_{-0.33} \pm 0.05$ and
 $\cala_{\kstarz\gamma}=-0.20 \pm 0.24 \pm 0.05$.
\end{abstract}


\maketitle

\tighten

{\renewcommand{\thefootnote}{\fnsymbol{footnote}}}
\setcounter{footnote}{0}

The radiative $b\to s\gamma$ penguin process is sensitive to physics
beyond the standard model (SM), and time-dependent $CP$ violation in
decays of the type $\bz\to f_{CP}\gamma$, where $f_{CP}$ is a $CP$
eigenstate, has drawn much theoretical and experimental interest
recently~\cite{Atwood:1997zr,Atwood:2004jj,Grinstein:2004uu,Grinstein:2005nu,Matsumori:2005ax,Ball:2006cv,Ushiroda:2005sb,Aubert:2005bu}.
Within the SM, the
photon emitted from a $\bz$ ($\bzb$) meson is predominantly right-handed
(left-handed). A flip of photon polarization is suppressed by the quark
mass ratio $2m_s/m_b$~\cite{Atwood:1997zr}.  Hence, for $f_{CP} =
\kstarz\ (\to\ks\piz)$ the SM predicts a small time-dependent $CP$
asymmetry, which arises from the interference between decay amplitudes with
and without $\bz$-$\bzb$ mixing.
The same suppression is expected for final states of the type $\bz\to
P^0Q^0\gamma$~\cite{Atwood:2004jj}, where $P^0$ and $Q^0$ are any $C$
eigenstate spin-0 neutral particle (e.g. $P^0 = \ks$ and $Q^0 =
\piz$).
However, there are estimates predicting an enhancement of the asymmetry
up to 0.1 due to strong interactions~\cite{Grinstein:2004uu,Grinstein:2005nu}.
For the case of $\bz\to\kstarz\ (\to\ks\piz)\gamma$, explicit
computations support a small
asymmetry, $\cals = -(3.5\pm 1.7)\times 10^{-2}$~\cite{Matsumori:2005ax}
or $\cals = -(2.2\pm 1.2{}^{+0}_{-1.0})\times 10^{-2}$~\cite{Ball:2006cv} in the SM.
A significant deviation from the small SM expectation could indicate new
physics.

Since the
time-dependent $CP$ asymmetry
is not expected to change significantly as a
function of $\ks\piz$ invariant mass
($M_{\ks\piz}$)~\cite{Grinstein:2005nu}, we perform two measurements:
one for $\bz\to\kstarz(\to\ks\piz)\gamma$~\cite{bib:CC}
by requiring $M_{\ks\piz}$ to
lie in the range $0.8\GeVcc<M_{\ks\piz}<1.0\GeVcc$, and the other for
the full range of $M_{\ks\piz}$ below $1.8\GeVcc$. For simplicity, we
refer to these two analyses as $\kstarz\gamma$ and $\ks\piz\gamma$,
respectively. The measurement for $\kstarz\gamma$ is theoretically
cleaner than the measurement for $\ks\piz\gamma$, but the latter has
more statistical power.  Similar measurements have been previously
reported by both Belle~\cite{Ushiroda:2005sb} and
\BaBar~\cite{Aubert:2005bu} based on $\NBBofour$ and $232\times 10^6$
$B\bbar$ pairs, respectively. In this Communication, we update the
measurements of $CP$ parameters for $\bz\to\kstarz\gamma$ and
$\bz\to\ks\piz\gamma$ based on a data sample that contains $\NBBosix$
$B\bbar$ pairs.

At the KEKB asymmetric-energy $e^+e^-$ (3.5 on 8.0$\GeV$)
collider~\cite{bib:KEKB}, the $\Upsilon(4S)$ is produced with a Lorentz
boost of $\beta\gamma=0.425$ along the $z$ axis, which is defined as the
direction antiparallel
to the $e^+$ beam direction.
In the decay chain $\Upsilon(4S)\to \bz\bzb \to \fCP \ftag$, where one
of the $B$ mesons decays at time $\tCP$ to a final state $\fCP$, which
is our signal mode, and the other decays at time $\ttag$ to a final
state $\ftag$ that distinguishes between $B^0$ and $\bzb$, the decay
rate has a time dependence given by
\begin{eqnarray}
\label{eq:psig}
{\cal P}(\Dt) = 
\frac{e^{-|\Dt|/{\taubz}}}{4{\taubz}}
\biggl\{1 &+& \fq
\Bigl[ \cals\sin(\dmd\Dt) \nonumber\\
   &+& \cala\cos(\dmd\Dt)
\Bigr]
\biggr\}.
\end{eqnarray}
Here $\cals$ and $\cala$ are $CP$-violation parameters, $\taubz$ is the
$B^0$ lifetime, $\dmd$ is the mass difference between the two $B^0$ mass
eigenstates, $\Dt$ is the time difference $\tCP - \ttag$, and the
$b$-flavor charge $\fq$ = +1 ($-1$) when the tagging $B$ meson is a
$B^0$ ($\bzb$).
Since the $B^0$ and $\bzb$ mesons are approximately at 
rest in the $\Upsilon(4S)$ center-of-mass system (c.m.s.),
$\Dt$ can be determined from the displacement in $z$ 
between the $\fCP$ and $\ftag$ decay vertices:
$\Delta t \simeq (\zCP - \ztag)/(\beta\gamma c)
 \equiv \Delta z/(\beta\gamma c)$.

The Belle detector is a large-solid-angle magnetic
spectrometer that
consists of a silicon vertex detector (SVD),
a 50-layer central drift chamber, an array of
aerogel threshold \v{C}erenkov counters, 
a barrel-like arrangement of time-of-flight
scintillation counters, and an electromagnetic calorimeter (ECL)
comprised of CsI(Tl) crystals located inside 
a superconducting solenoid coil that provides a 1.5~T
magnetic field.  An iron flux-return located outside of
the coil is instrumented to detect $K_L^0$ mesons and to identify
muons.  The detector
is described in detail elsewhere~\cite{Belle}.
Two inner detector configurations were used. A 2.0\,cm radius beampipe
and a three layer silicon vertex detector (SVD1) 
were 
used for the first sample
of $\NBBsvdI$ $B\bbar$ pairs, while a 1.5\,cm radius beampipe, a four layer
silicon detector (SVD2) and a small-cell inner drift chamber were used to record
the remaining $\NBBsvdII$ $B\bbar$ pairs~\cite{SVD2}.


For high energy prompt photons, we select the cluster in the ECL
with the highest energy in the c.m.s. from clusters that have no
associated charged track.
We require $1.4\GeV < \egcms < 3.4\GeV$.
For the selected photon, we also require $E_9/E_{25}>0.95$, where
$E_9/E_{25}$ is the ratio of energies summed in $3\times 3$ and $5\times
5$ arrays of CsI(Tl) crystals around the center of the shower.
In order to reduce the background from $\piz$ or $\eta$ decays, photons
from candidate $\piz\to\gamma\gamma$ or $\eta\to\gamma\gamma$ decays are
rejected using a likelihood described in detail
elsewhere~\cite{Koppenburg:2004fz}.
The polar angle of the photon direction in the laboratory frame is
restricted
to the barrel region of the ECL ($33^\circ < \theta_\gamma <
128^\circ$) for SVD1 data, but is extended to
the end-cap regions ($17^\circ < \theta_\gamma < 150^\circ$)
for SVD2 data due to the reduced material in front of the ECL.

Neutral kaons ($\ks$) are reconstructed from two oppositely charged pions
that have an invariant mass within $\pm 6\MeVcc$ ($2\sigma$) of the $\ks$
mass~\cite{bib:PDG06}.
The $\pip\pim$ vertex is required to be displaced from
the interaction point (IP) by a minimum transverse distance of 0.22\,cm for
high momentum ($>1.5\GeVc$) candidates and 0.08\,cm for
those with momentum less than $1.5\GeVc$. 
The direction of the pion pair momentum must agree with
the direction defined by the IP and the two pion vertex point
within 0.03 rad for high-momentum candidates, and within 0.1
rad for the remaining candidates.
Neutral pions ($\piz$) are formed from two photons with
an
invariant mass within $\pm 16\MeVcc$ ($3\sigma$) of the $\piz$ mass.
The photon momenta are then recalculated with a $\piz$ mass constraint.
We then require the momentum of $\piz$ candidates in the c.m.s. to
be greater than $0.3\GeVc$.  The $\ks\piz$ invariant mass,
$M_{\ks\piz}$, is required to be less than $1.8\GeVcc$.

We also reconstruct $\bz\to\kp\pim\gamma$ and $\bp\to\ks\pip\gamma$
candidates as control samples in a similar way.
Charged tracks other than those from $\ks$ are required to originate
from the IP (within 5\,cm in $z$ and 1.4\,cm in $r$-$\phi$);
the transverse momentum ($p_t$) is required to be greater than
$0.1\GeVc$.
In the $\bp\to\ks\pip\gamma$ sample, we
also require that the $\pip$ candidate
is not positively identified as
any other particle species ($\kp,\ p^+,\ e^+$ and $\mu^+$).
In the $\bz\to\kp\pim\gamma$ sample, the $\kp$ candidate is selected
from charged tracks identified as kaons, and the $\pim$ candidate from
the rest of the tracks.

The $\bz\to\ks\piz\gamma$ and $\bp\to\ks\pip\gamma$ modes are
reconstructed simultaneously and a single candidate is selected from
possible multiple candidates amongst the two modes in order to reduce
the cross-feed background from $\bp\to\ks\pip\gamma$ in
$\bz\to\ks\piz\gamma$, where a $\piz$ is selected instead of the $\pip$.
The best candidate selection is based on a
likelihood ratio $\LR$ calculated from likelihood
variables for signal ($\mathcal{L}_{sig}$) and background
($\mathcal{L}_{bkg}$) as $\LR \equiv
\mathcal{L}_{sig}/(\mathcal{L}_{sig}+\mathcal{L}_{bkg})$, where
the likelihood variables are
obtained from a Fisher discriminant $\calf$~\cite{Fisher}, which uses
the modified Fox-Wolfram moments~\cite{Abe:2003yy} as
discriminating variables. Hereafter, we denote the likelihood ratio with
the likelihood variable name in parentheses, as $\LR$($\calf$) in
this case. We select the candidate that has the largest $\LR$($\calf$).
We form two kinematic variables: the energy difference
$\dE\equiv E_B^{\rm c.m.s.}-E_{\rm beam}^{\rm c.m.s.}$ and the beam-energy
constrained mass $\mb\equiv\sqrt{(E_{\rm beam}^{\rm c.m.s.})^2-(p_B^{\rm
c.m.s.})^2}$, where $E_{\rm beam}^{\rm c.m.s.}$ is the beam energy,
and $E_B^{\rm c.m.s.}$ and $p_B^{\rm c.m.s.}$ are the energy and the
momentum of the candidate in the c.m.s.
The signal region in $\dE$ and $\mb$, which is used for the
measurements of $CP$-violating parameters,
is defined as $-0.2\GeV < \dE < 0.1\GeV$ and
$5.27\GeVcc < \mb < 5.29\GeVcc$. In order to determine the
$\dE$-$\mb$ dependent signal fraction, a larger fitting region,
$-0.3\GeV < \dE < 0.5\GeV$ and $5.2\GeVcc < \mb$ is used.

In order to suppress the background contribution from continuum light
quark pair production processes ($e^+e^-\to q\,\bar{q}$ with $q =
u,d,s,c$), which we hereafter refer to as $\qq$, we form another
likelihood ratio $\LR$($\calf$,$\cos\theta_B$,$\cos\theta_H$)
by combining $\calf$ with $\cos\theta_B$ and $\cos\theta_H$,
where $\theta_B$ is the polar angle of the $B$ meson candidate momentum
in the c.m.s. and $\theta_H$ is the helicity angle defined as
the kaon momentum direction with respect to the opposite of the $B$
momentum in the $K$-$\pi$ rest frame.
The helicity distributions for signal and background are determined from
the $\bz\to\kp\pim\gamma$ sample for three mass regions:
$0.8\GeVcc < M_{K\pi} < 1.0\GeVcc$ (MR1),
$1.3\GeVcc < M_{K\pi} < 1.55\GeVcc$ (MR2),
and the remaining range up to $1.8\GeVcc$ (MR3).
The difference of background distributions in $\bz\to\kp\pim\gamma$ and
$\bz\to\ks\piz\gamma$ decay modes is corrected using sideband data samples.
In addition, a helicity dependent efficiency correction is applied.
Specific $\LR$($\calf$,$\cos\theta_B$,$\cos\theta_H$) selection criteria are
applied depending on both the mass region and the flavor tagging
information.
Background contributions from $B$ decays, which are considerably smaller
than $\qq$, are dominated by cross-feed from other radiative $B$ decays.


The $b$-flavor of the accompanying $B$ meson is identified from
inclusive properties of particles that are not associated with the
reconstructed signal decay.  The algorithm for flavor tagging is
described in detail elsewhere~\cite{bib:fbtg_nim}.  We use two
parameters, $\fq$ defined in Eq.~(\ref{eq:psig}) and $r$, to represent
the tagging information.  The parameter $r$ is an event-by-event
flavor-tagging quality factor that ranges from 0 to 1: $r=0$ when there
is no flavor discrimination and $r=1$ when the flavor assignment is
unambiguous. The value of $r$ is determined by using Monte Carlo (MC)
and is only used to sort data
into seven $r$ intervals. Events with $r>0.1$ are sorted into six $r$
intervals.  The wrong tag fraction $w$ and the difference $\Delta w$ in
$w$ between the $\bz$ and $\bzb$ decays are determined for each of the
six $r$ intervals from high-statistics control samples of semi-leptonic
and hadronic $b\to c$ decays.
If $r$ is less than or equal to 0.1, we set $w$ to 0.5, and therefore
the accompanying $B$ meson does not provide tagging information in this
case.


The vertex position of the signal-side decay of $\bz\to\ks\piz\gamma$
and $\bp\to\ks\pip\gamma$ is reconstructed from the $\ks$ trajectory
with a constraint on the IP; the IP profile ($\sigma_x\simeq 100\rm\,\mu
m$, $\sigma_y\simeq 5\rm\,\mu m$) is smeared by the finite $B$
flight length in the plane perpendicular to the $z$ axis.
The $\ks$ vertex is displaced from the $B$ vertex and often lies outside
of the SVD. In this case, the vertex resolution is not good enough for a
time-dependent $CP$ asymmetry measurement. Therefore, both pions from
the $\ks$ decay are required to have enough hits in the SVD: at least
one layer with hits on both the $z$ and $r$-$\phi$ sides and at least one
additional hit in the $z$ side of the other layers for SVD1, and at
least two layers with hits on both sides for SVD2.
The other (tag-side) $B$ vertex is determined from well reconstructed
tracks that are not assigned to the signal side.
A constraint to the IP profile is also imposed.

After all the selections
are 
applied, we obtain $\NkspizgammaCandInFit$
candidates in the $\dE$-$\mb$ fit region,
of which
$\NkspizgammaCandInBox$ are in the signal box. We perform an unbinned
maximum likelihood (UML) fit to
the $\dE$-$\mb$ distribution in order to resolve signal, $B\bbar$
background and $\qq$ background components.
The signal probability density function (PDF) is obtained from MC.
We use a two-dimensional histogram of MC simulated data,
for which the peak position and the width are corrected
to account for differences between data and simulation
using the $\bz\to\kp\pim\gamma$ control sample.
The two-dimensional PDF for the $B\bbar$ background,
which populates more the lower $\dE$ region, is obtained from MC.
For $\qq$ background, we use the product of two one-dimensional PDFs:
the ARGUS parameterization~\cite{bib:ARGUS} for $\mb$ and a 
second order polynomial for $\dE$.
Five parameters, which are the signal fraction, the $B\bbar$ background
fraction, ARGUS shape parameter $\alpha$,
and two polynomial coefficients ($c_1$, $c_2$),
are the free parameters in the fit.

\begin{figure}
 \resizebox{0.46\columnwidth}{!}{\includegraphics{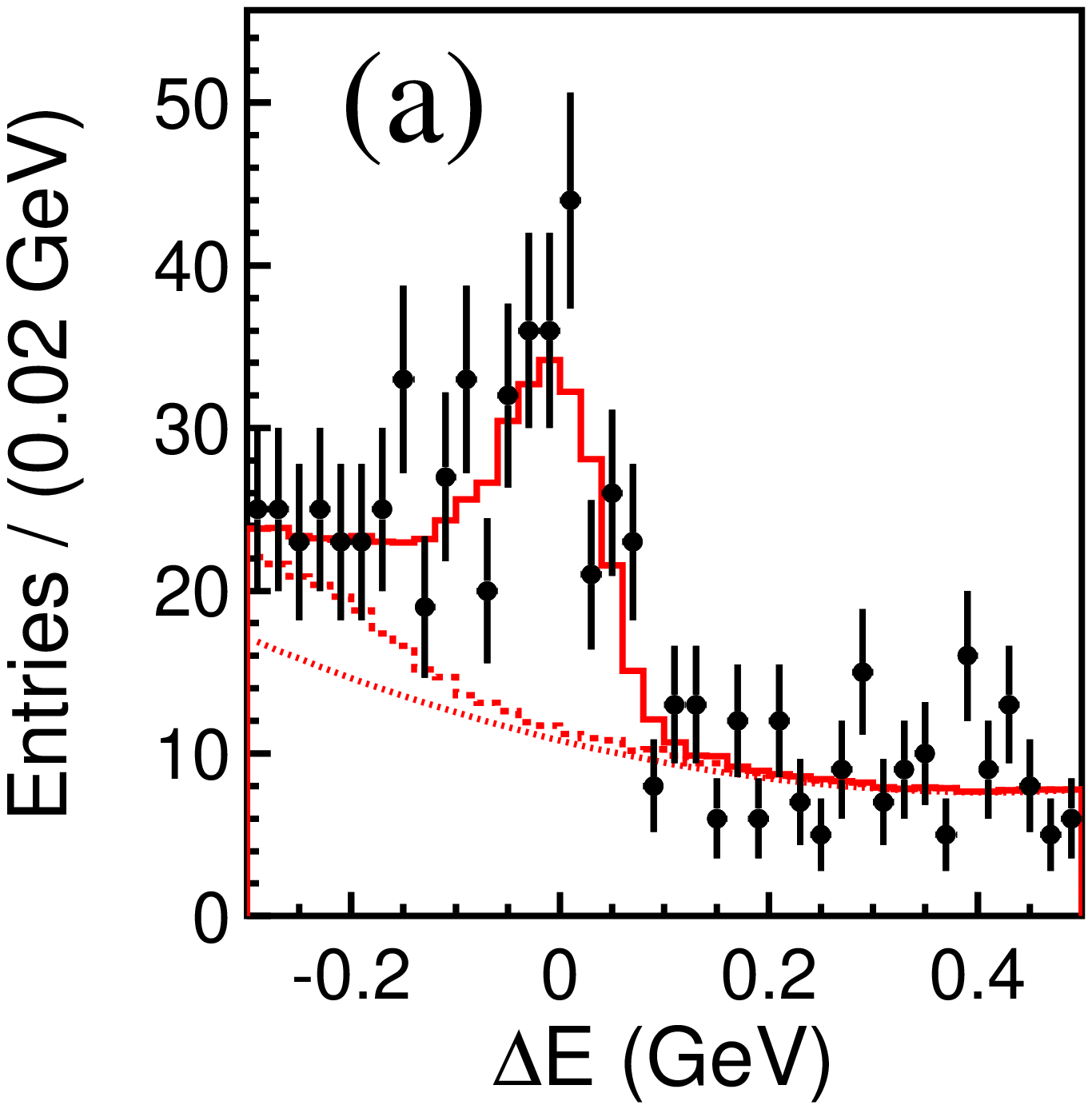}}
 \resizebox{0.46\columnwidth}{!}{\includegraphics{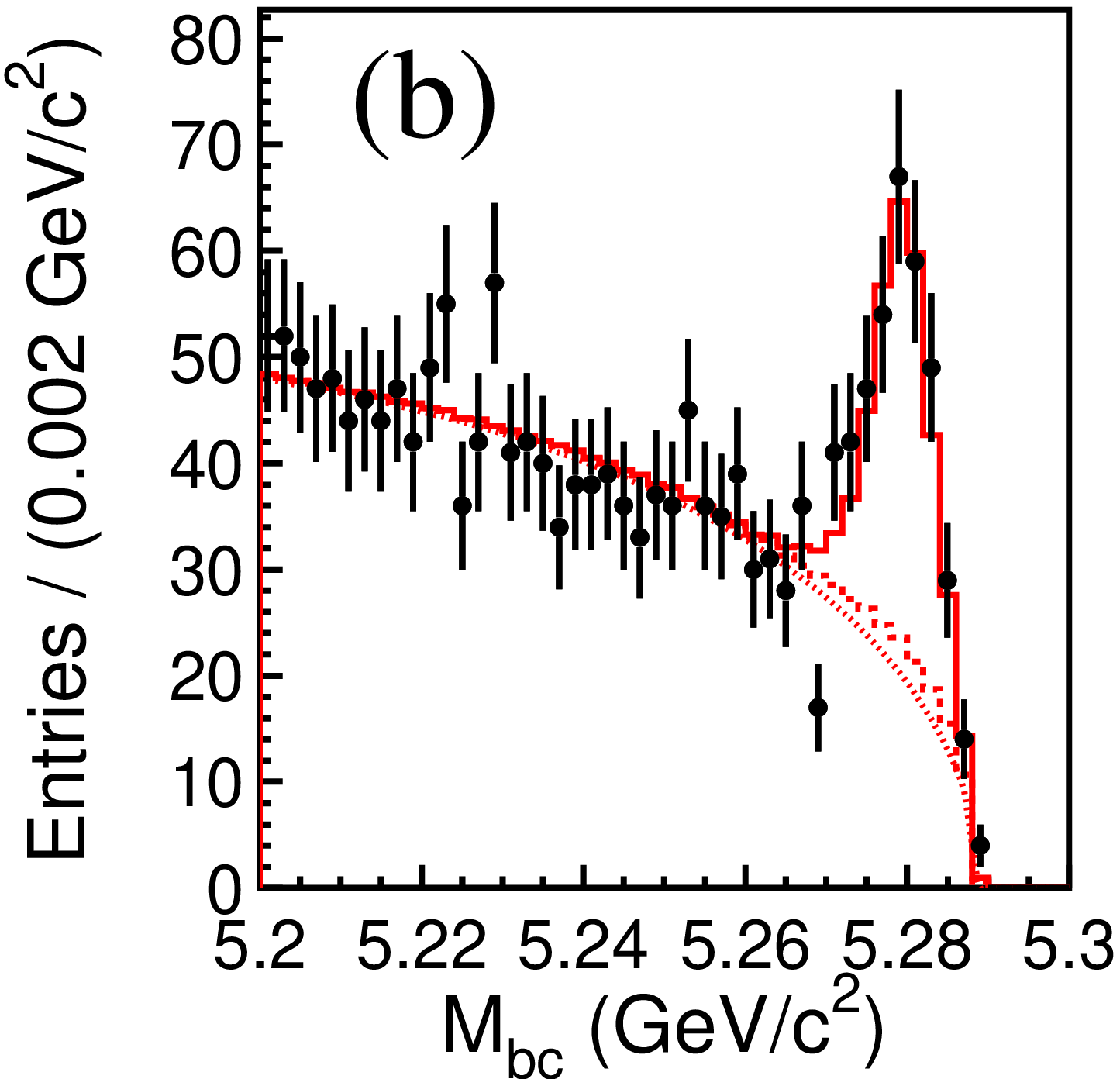}}\\
 \caption{
 (a) $\dE$ distribution within the $\mb$ signal slice
 and (b) $\mb$ distribution within the $\dE$ signal slice for the
 whole $M_{\ks\piz}$ region. Points with error bars are measured data.
 The solid curves show the fit results. The dotted curves show the
 $\qq$ background contributions, while the dashed curves show the sum
 of $\qq$ and $B\bbar$ background contributions.
 }
 \label{fig:mbcde}
\end{figure}

We first fit the entire $M_{\ks\piz}$ region, and then fit the MR1,
MR2 and MR3 mass regions separately with the three background shape
parameters ($\alpha$, $c_1$ and $c_2$) fixed at the values obtained from
the fit to the full range.
Figure~\ref{fig:mbcde} shows a $\dE$ ($\mb$) projection of the
fit result in the $\mb$ ($\dE$) signal slice for the
entire $M_{\ks\piz}$ region.
For the MR1, MR2, and MR3 samples, we find
$\syme{\NsignMRI  }{\NsignMRIe  }$,
$\syme{\NsignMRII }{\NsignMRIIe }$,
and $\syme{\NsignMRIII}{\NsignMRIIIe}$ signal events, respectively,
with signal-to-background ratios (S/N) of
$\SNnMRI$, $\SNnMRII$, and $\SNnMRIII$.
The level of $B\bbar$ background is only
$\syme{\NrarebnMRI  }{\NrarebnMRIe  }$,
$\syme{\NrarebnMRII }{\NrarebnMRIIe }$,
and $\syme{\NrarebnMRIII}{\NrarebnMRIIIe}$ $B\bbar$ events, respectively.
As expected, in the $\kstarz$ mass region the S/N ratio is high.

We determine $\cals$ and $\cala$ from an UML fit
to the observed $\Dt$ distribution. The
PDF expected for the signal distribution, ${\cal P}_{\rm
sig}(\Dt;\cals,\cala,\fq,w,\Delta w)$, is given by the time-dependent
decay rate [Eq.~(\ref{eq:psig})], modified to 
incorporate the effect of incorrect flavor assignment;
the parameters $\tau_\bz$ and $\dmd$ are fixed at
their world-average values~\cite{bib:PDG06}.
The distribution is then convolved with the
proper-time interval resolution function $\Rsig$, which takes into
account the finite vertex resolution.
The parameterization of $\Rsig$ is the same as that in the previous
measurement~\cite{Ushiroda:2005sb}, while the parameter values
are updated for the whole dataset.
The PDF for $B\bbar$ background events (${\cal P}_{B\bbar}$) is modeled
in the same way as signal, but with different lifetime and $CP$-violating
parameters, while  the resolution function $\Rbb$ is the same as $\Rsig$.
The effective lifetime of $B\bbar$ background is obtained from
a fit to the MC sample; the result is
$\rarebzlife\,$ps.
The background is assumed to have no $CP$ asymmetry.  Since 20\% of the
$B\bbar$ background are from non-radiative $\bz$ decay, assigning the
maximum asymmetries to this component, possible $CP$
asymmetries in the background ($\cals = \pm 0.2$ and $\cala = \pm 0.2$)
are taken into account in the systematic error.

The PDF for $\qq$ background events, ${\cal P}_{\qq}$, is modeled as a
sum of exponential and prompt components, and is convolved with a
double Gaussian which represents the resolution function $\Rbkg$.
All parameters in ${\cal P}_{\qq}$ and $\Rbkg$ are
determined by a fit to the $\Dt$ distribution of a background-enhanced
sample in the $\dE$-$\mb$ sideband region.

For each event, the following likelihood function is
evaluated:
\begin{equation}
  \begin{split}
   P_i
   =& (1-\fol)\int_{-\infty}^{+\infty} \biggl[
   \fsig{\cal P}_{\rm sig}(\Dt')\Rsig (\Dt_i-\Dt') \\
   &+\fbb{\cal P}_{B\bbar}(\Dt')\Rbb (\Dt_i-\Dt') \\
   &+(1-\fsig-\fbb){\cal P}_{\qq}(\Dt')\Rbkg (\Dt_i-\Dt')\biggr]
   d(\Dt')  \\
   &+\fol P_{\rm ol}(\Dt_i),
   \label{eq:likelihood}
  \end{split}
\end{equation}
where $P_{\rm ol}$ is a Gaussian function that represents a small
outlier component with fraction $\fol$~\cite{bib:resol}.
The probability of signal ($\fsig$) and background ($\fbb$) are
calculated on an event-by-event
basis using the results of the two-dimensional $\dE$-$\mb$ fit,
and are then multiplied
by a factor that depends on the flavor tagging bin $r$.
The $r$ distributions of signal and $\qq$ background events are
estimated by repeating the $\dE$-$\mb$ fit procedure for each $r$
interval with the three background shape parameters fixed to the full range result.
The $B\bbar$ background distribution is estimated from MC
since the number of $B\bbar$ background events in data is limited.

\begin{figure}
 \resizebox{0.46\columnwidth}{!}{\includegraphics{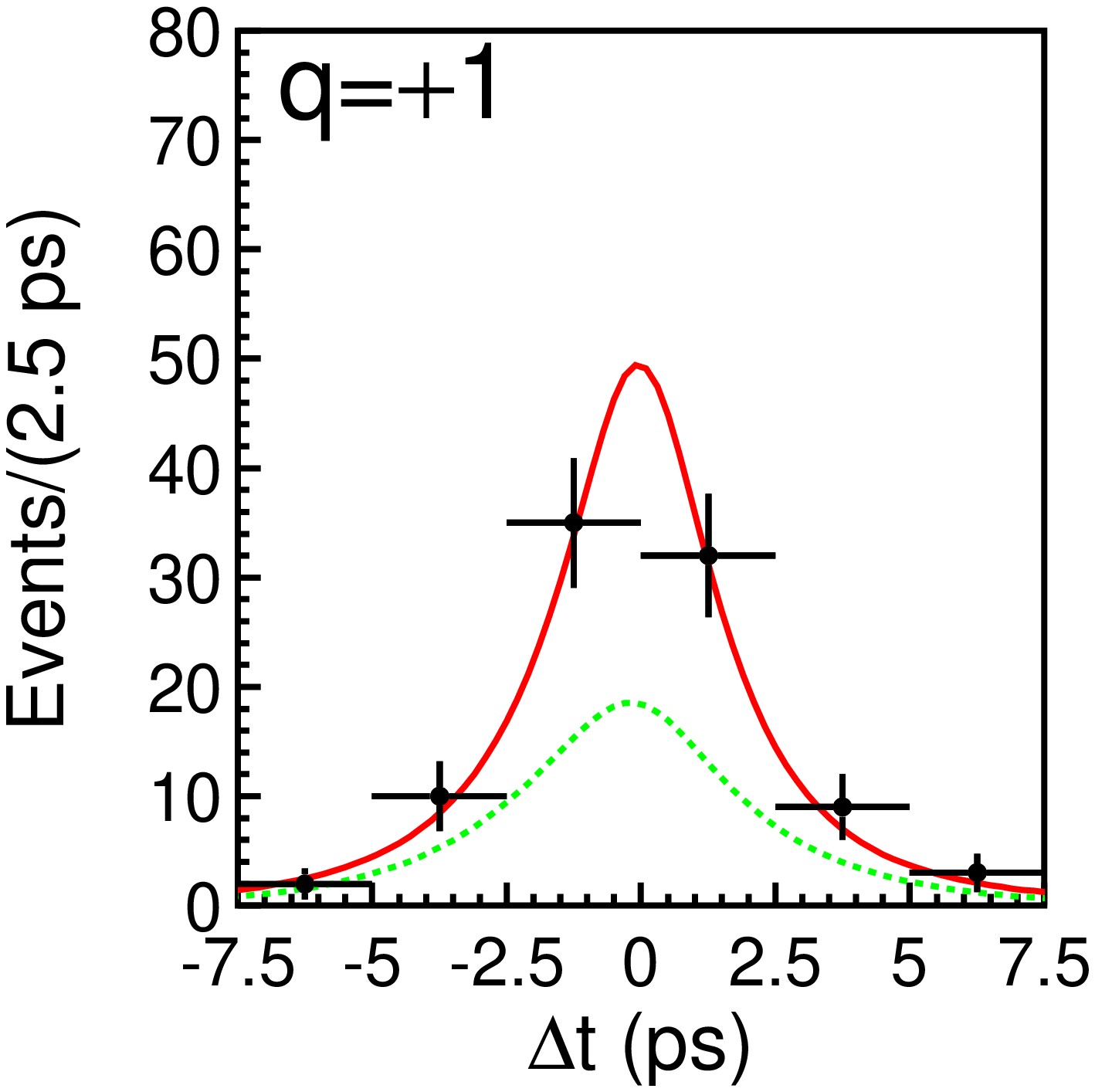}}
 \resizebox{0.46\columnwidth}{!}{\includegraphics{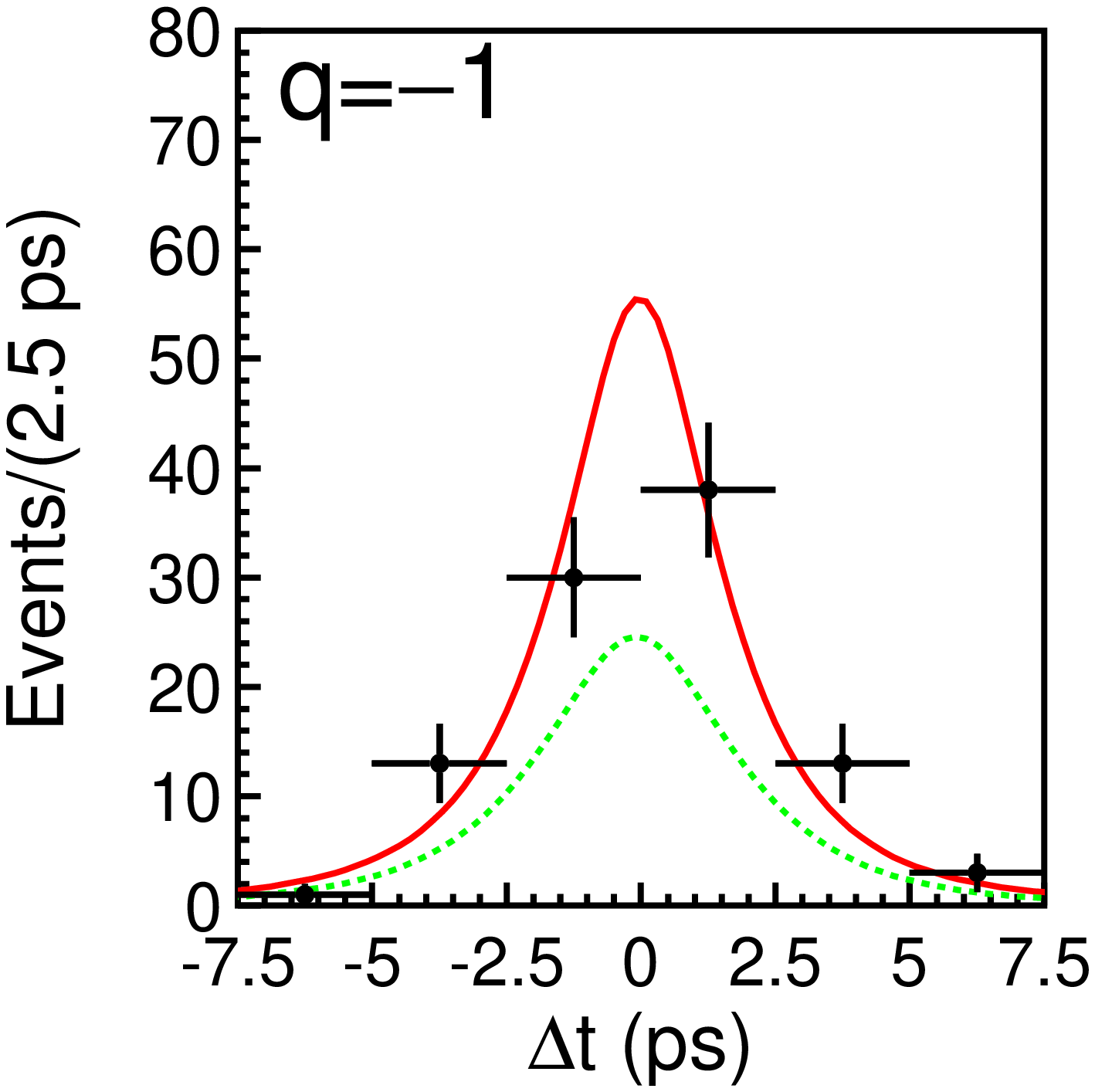}}
 \resizebox{0.46\columnwidth}{!}{\includegraphics{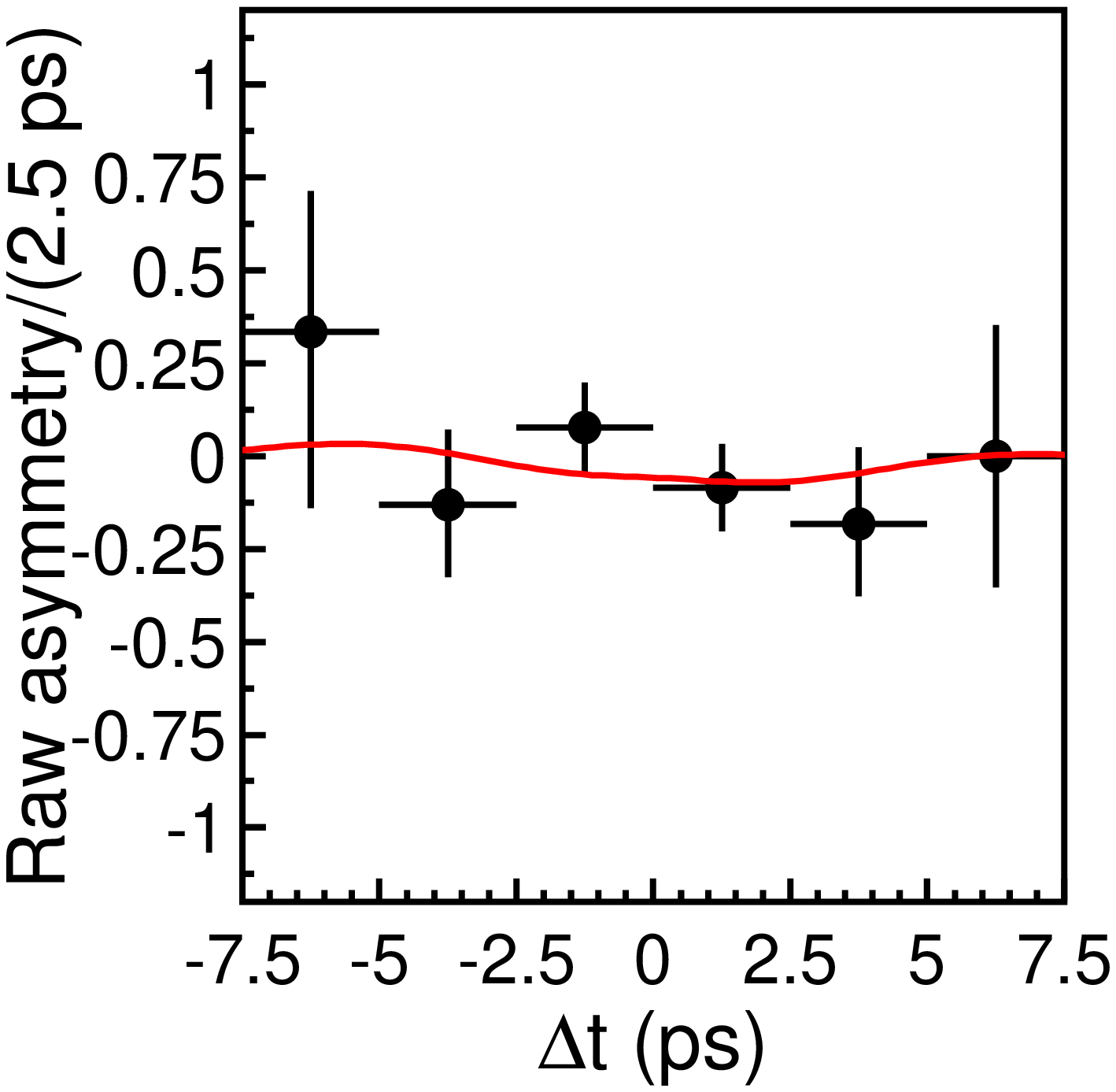}}
\caption{
 (Top) Proper time distributions for $\bz\to\ks\piz\gamma$
 for $q=+1$ (left) and $q=-1$ (right) with $0.5 < r \le 1.0$.
 The solid curve shows the total and dashed curve shows the signal component.
 (Bottom) Asymmetry in each $\Dt$ bin with $0.5 < r \le 1.0$.
 The solid curve shows the result of the UML fit.
 }
\label{fig:asym}
\end{figure}

The only free parameters in the $CP$ fit to $\bz\to\ks\piz\gamma$
are $\cals_{\ks\piz\gamma}$ and $\cala_{\ks\piz\gamma}$, which are
determined by maximizing the likelihood function $L =
\prod_iP_i(\Dt_i;\cals,\cala)$, where the product is over all events.
We obtain
\begin{eqnarray}
 \cals_{\ks\piz\gamma} &=& \SkspizgmResultSS, \\
 \cala_{\ks\piz\gamma} &=& \AkspizgmResultSS,
\end{eqnarray}
where the systematic errors are obtained as discussed below.
We define the raw asymmetry in each $\Dt$ bin by
$(N_{q=+1}-N_{q=-1})/(N_{q=+1}+N_{q=-1})$, where $N_{q=+1~(-1)}$ is the
number of observed candidates with $q=+1~(-1)$.  Figure~\ref{fig:asym}
shows the $\Dt$ distributions of the events with $0.5 < r \le 1.0$
for $q=+1$ and $q=-1$ and the raw asymmetry.

We perform the following fits to confirm the validity of our procedure:
a $\bp$ lifetime fit for the $\bp\to\ks\pip\gamma$ sample gives
$\controllife\,$ps,
which is consistent with the nominal $\bp$ lifetime~\cite{bib:PDG06};
a $\bz$ lifetime fit for the $\bz\to\ks\piz\gamma$ sample gives
$\signallife\,$ps,
which is consistent with the nominal $\bz$ lifetime~\cite{bib:PDG06};
and a $CP$ asymmetry fit for the $\bp\to\ks\pip\gamma$ sample gives
an asymmetry consistent with zero ($\cals = \controlSin$, $\cala =
\controlCos$).
This is expected since the charged decay is completely flavor specific
regardless of the photon polarization.
A fit to MR1 data gives
$\cals_{\kstarz\gamma} = \SMRIResultSS$,
$\cala_{\kstarz\gamma} = \AMRIResultSS$.
A fit to non-MR1 data gives $\cals = \SnonResultSS$ and $\cala =
\AnonResultSS$. These results are consistent with those from the full
$M_{\ks\piz}$ sample.

We evaluate systematic uncertainties in the following
categories by fitting the data with
each fixed parameter shifted by its error:
uncertainties from physics parameters such as $\dmd$, $\tau_\bz$,
effective lifetime and $CP$ asymmetry of $B\bbar$ background,
uncertainties in the knowledge of $\qq$ background $\Delta t$ PDF,
uncertainties in the flavor tagging,
uncertainties in the signal and background fractions,
uncertainties in the resolution functions.
A possible bias in the fit is checked by fitting a large MC sample. The fit
result is consistent with the input value within the statistical error.
We quote this statistical error as the possible fit bias.
Uncertainty due to the vertex reconstruction is estimated by using the
$\bz\to J/\psi\ks$ control sample, where the tracks from $J/\psi$ are
ignored for the study of the signal side vertex reconstruction using
the $\ks$ trajectory.
The effect of SVD misalignment is estimated by artificially displacing the
SVD sensors in a random manner; the standard deviation of these shifts and
rotations are $15\,\mu$m and $0.15\,$mrad, respectively.
Effects of tag-side interference~\cite{Long:2003wq} are estimated using
$B\to D^\ast\ell\nu$.
The dominant systematic contributions on $\cals$ are
from the uncertainties in the signal and background fraction and the
resolution function.
The systematic error on $\cala$ is dominated by the tag-side interference.
All these contributions to the
systematic errors are
summed in quadrature to give
$0.07$, $0.05$ and $0.29$ for $\cals$ and
$0.06$, $0.05$ and $0.13$ for $\cala$
for the full $M_{\ks\piz}$, MR1 and non-MR1 samples, respectively.

Ensemble tests are carried out with MC pseudo-experiments using
the values of $\cals$
and $\cala$ obtained by the fit as the input parameters.
From 10,000 pseudo-experiments, we find that the
statistical errors obtained in our measurement
are within expectations.

In summary, we have performed a measurement of the time-dependent $CP$
asymmetry in the decay $\bz\to\ks\piz\gamma$ with 
$\ks\piz$ invariant mass up to $1.8\GeVcc$, based on a sample of
$\NBBosix$ $B\bbar$ pairs.
We obtain $CP$-violation parameters
$\cals_{\ks\piz\gamma}=\SkspizgmResultSS$ and
$\cala_{\ks\piz\gamma}=\AkspizgmResultSS$ for the full $\ks\piz$
invariant mass region, and
$\cals_{\kstarz\gamma}=\SMRIResultSS$ and
$\cala_{\kstarz\gamma}=\AMRIResultSS$ for the mass region around
$\kstarz(892)$.
This measurement supersedes our previous
measurement~\cite{Ushiroda:2005sb}.
With the present statistics,
we do not find any significant $CP$ asymmetry and therefore no
indication of new physics from right-handed currents.

We thank the KEKB group for excellent operation of the
accelerator, the KEK cryogenics group for efficient solenoid
operations, and the KEK computer group and
the NII for valuable computing and Super-SINET network
support.  We acknowledge support from MEXT and JSPS (Japan);
ARC and DEST (Australia); NSFC and KIP of CAS (China); 
DST (India); MOEHRD, KOSEF and KRF (Korea); 
KBN (Poland); MIST (Russia); ARRS (Slovenia); SNSF (Switzerland); 
NSC and MOE (Taiwan); and DOE (USA).

\end{document}